\documentclass[twocolumn]{aastex631}

\usepackage{color}
\usepackage{soul}

\begin{document}

\title{Steep Balmer decrement in weak AGNs may be not caused by dust extinction: clues from low-luminosity AGNs and changing-look AGNs}

\correspondingauthor{Qingwen Wu}
\email{qwwu@hust.edu.cn}

\author[0000-0002-2581-8154]{Jiancheng Wu}
\affiliation{Department of Astronomy, School of Physics, Huazhong University of Science and  Technology, Luoyu Road 1037, Wuhan, China}

\author[0000-0003-4773-4987]{Qingwen Wu$^*$}
\affiliation{Department of Astronomy, School of Physics, Huazhong University of Science and  Technology, Luoyu Road 1037, Wuhan, China}

\author[0000-0002-8064-0547]{Hanrui Xue}
\affiliation{Department of Astronomy, School of Physics, Huazhong University of Science and  Technology, Luoyu Road 1037, Wuhan, China}

\author[0000-0003-3440-1526]{Weihua Lei}
\affiliation{Department of Astronomy, School of Physics, Huazhong University of Science and  Technology, Luoyu Road 1037, Wuhan, China}

\author[0000-0001-8879-368X]{Bing Lyu}
\affiliation{
Kavli Institute for Astronomy and Astrophysics, Peking University, Beijing 100871, Peoples Republic of China}

\begin{abstract}
The hydrogen Balmer decrement (e.g., $\rm H\alpha/H\beta$) is widely adopted as an indicator of the internal reddening of active galactic nuclei (AGNs). This is challenged by some low-luminosity AGNs (LLAGNs) and changing-look AGNs (CLAGNs), which have steep Balmer decrement but without strong evidence for absorption. We compile a sample of normal AGNs and CLAGNs with a wider distribution of bolometric Eddington ratio ($\lambda_{\rm Edd}=L_{\rm bol}/L_{\rm Edd}$) and find a strong negative correlation between $\rm H\alpha/H\beta$ and $\lambda_{\rm Edd}$, which suggests that the Balmer decrement is also accretion-rate dependent. We further explore the Balmer decrement based on the photoionization model using the Cloudy code by considering spectral energy distribution (SED) from the accretion disk with different accretion rates (e.g., disk/corona and truncated disk at high and low Eddington ratios, respectively). Both the standard disk and truncated disk predict a negative correlation of $\rm H\alpha/H\beta-\lambda_{\rm Edd}$, where the relation is steeper in the case of the truncated disk. The negative correlations are also explored in two single CLAGNs. The measured negative correlation of $\rm H\alpha/H\beta$ -- $\lambda_{\rm Edd}$ is mainly caused by the lower responsivity $({\rm dlog}L_{\rm line}/{\rm dlog}L_{\rm cont})$ in $\rm H\alpha$ relative to that in $\rm H\beta$, due to the larger optical depth in the former. We propose that the steep Balmer decrements in low-Eddington-ratio AGNs (e.g., some Seyferts 1.5-1.9 and CLAGNs) are not simply caused by absorption but mainly caused by the relatively low flux of ionizing photons.



\end{abstract}

\keywords{Active galactic nuclei (16), Seyfert galaxies (1447), Accretion (14), Supermassive black holes (1663), Line intensities (2084)}

\section{Introduction} \label{sec:intro}

Active galactic nuclei (AGN) are powered by the accretion of matter onto a supermassive black hole (SMBH), which are the most luminous objects in our universe. The AGNs are empirically divided into type 1 and type 2 sources according to the presence of broad emission lines (e.g., Full Width at Half Maximum, $\rm FWHM\ge 1000 km/s$) or not. The broad emission lines are produced in the broad line region (BLR) that stay $\sim$pc scale from the central SMBH, and the narrow emission lines are emitted from the gas located at much larger distances (narrow line region, NLR). The type 1 and type 2 AGNs are unified based on the inclination of the accretion disk to the line of sight, where the BLR is obscured in type 2 AGNs by a putative dusty torus. The highly polarized components of broad emission lines as detected in some type 2 AGNs do support the above unification scenario \citep[e.g.,][]{Antonucci1993, Urry1995}. Intermediate-type AGNs (e.g., type 1.2, 1.5, 1.8, or 1.9) based on the relative flux of narrow and broad emission lines are also classified, which complicate the simple dichotomy \citep[e.g.,][]{Osterbrock1976,Osterbrock1981}. For example, the broad H$\beta$ component is weak or disappears in type 1.9 AGNs, where broad H$\alpha$ is still evident. A further challenge to the unification model comes from the so-called changing-look AGNs(CLAGNs), where a significant change in optical broad emission lines and/or strong variation of line-of-sight column densities are found in a single AGN within several years or several decades \citep[see a recent review,][]{Ricci2022}. The physical mechanism of CLAGNs is not well understood, and possible mechanisms include the variation of the obscuring material in clumpy torus \citep[e.g.,][]{Marin2013,Agsgonzlez2014,Rivers2015,Turner2018,Wang2019} or the variation of accretion rate \citep[e.g.,][]{Noda2018,Yang2021,Lyu2021,Lyu2022,Liu2022}.

The optically thick, geometrically thin standard accretion disk \citep[SSD,][]{Shakura1973} is believed to be presented in bright AGNs, where the big blue bump provides good evidence for this disk component. However, the SSD likely transitions to an optically thin, geometrically thick, advection-dominated accretion flow (ADAF) in the low luminosity AGNs \citep[LLAGNs,][]{Yuan2014}. Many LLAGNs show only a red ``bump", which possibly comes from the outer SSD \citep[e.g., the truncated disk model,][]{Ho2008}. Broad H$\alpha$ lines are observed while H$\beta$ and other high-ionization lines are weak or absent in many LLAGNs \citep[see more details in][]{Ho2008}. The possible physical reason may be the deficit of ionization photons in ADAF case \citep[e.g.,][]{Elitzur2009} or the BLR cannot be sustained due to the decrease of mass outflow rate of disk winds in such a low accretion state \citep[e.g.,][]{Elitzur2006}.

Balmer decrement has long been adopted as an indicator of the internal reddening of AGNs, which is measured from the strength of the hydrogen emission corresponding to different excitation states. It is normally assumed that the conditions (e.g., density, temperature, etc.) are fixed and high-ionization lines may be easily affected by the obscuring material since the optical depth is wavelength-dependent. Therefore, the deviation from the expected flux ratio of $\rm H\alpha/H\beta$ will indicate the amount of obscuration toward the line-emitting regions \citep[e.g.,][]{Ward1987,Gaskell2017,Osterbrock1989}. \cite{Gaskell2017} proposed that, with typical parameters of BLR, the intrinsic Balmer decrement in extremely blue AGNs is $\rm H\alpha/H\beta\sim2.7$ for broad emission lines, which is more or less consistent with a Baker-Menzel Case B value. However, the classical nebular theory of hydrogen line emission (Case B) is not expected to be applicable in the dense and optically thick conditions thought to exist in the line-emitting gas within the BLRs of AGN \cite[e.g.,][]{Netzer1975,Ferland1979_2,Korista2004,Ferland2020}. We also note that a Balmer line flux ratio measurement that happens to correspond to a prediction of Case B does not indicate the line-emitting gas conditions are those required by Case B: modestly optically thick in the Lyman lines, optically thin in the excited-state transition series, and spectrum completely dominated by recombination cascade. In general, there is no such one-to-one mapping of the hydrogen line flux ratio to physical line-emitting gas conditions. From the observations, the Balmer decrements were explored with several samples from Sloan Digital Sky Survey (SDSS). For example, \cite{LaMura2007} derived $\left \langle \rm H\alpha/H\beta \right \rangle=3.45\pm0.65$ based on 90 Seyfert 1 galaxies. \cite{Dong2008} selected 446 Seyfert 1 galaxies and found $\left \langle \rm H\alpha/H\beta \right \rangle=3.06$ with standard deviation 0.03 dex. \cite{Lu2019} found  $\left \langle \rm H\alpha/H\beta \right \rangle=3.16$ with a standard deviation of 0.07 dex for 554 low-redshift AGNs. 
It is still unknown whether the higher observational ratio of $\rm H\alpha/H\beta$ compared to the theoretical prediction is intrinsic or caused by the internal reddening of AGNs.

Furthermore, in some local Seyferts, the ratio of $\rm H\alpha/H\beta$ ranges from 4 to 7 \citep[e.g.,][]{Schnorr2016}, which is even larger in some type 1.8/1.9 AGNs \citep{Crenshaw1988,Osterbrock1976}. The $\rm H\alpha/H\beta$ in some optical CLAGNs showing the appearance and disappearance of broad emission lines can reach up to 10 or even higher with a decrease of luminosities (\citealp[e.g., $\rm H\alpha/H\beta=8-15$ at 150 days after the transient in 1ES 1927+654,][]{Trakhtenbrot2019}; $\rm H\alpha/H\beta=8-10$ in NGC 2992, \citealp{Guolo2021};  $\rm H\alpha/H\beta=4-6$ in NGC 3516, \citealp{Shapovalova2019}; $\rm H\alpha/H\beta=4-9$ in Mrk 1018, \citealp{Cohen1986,McElroy2016}). The effects of collisional excitation and/or dust extinction are suggested to be the reason for such a large deviation of the Balmer decrement \cite[][]{Gaskell_Ferland1984,Ost_Ferland2006,Shapovalova2019}. It should be noted that the absorption is always low in some CLAGNs \citep[e.g., Mrk 1018,][]{LaMassa2017} and some Seyfert 1.8/1.9 galaxies, \cite[e.g.,][]{Barcons2003}. Therefore, the absorption scenario cannot fully explain the steep Balmer decrement in these AGNs.

The broad emission lines in AGNs are sensitive to the spectral energy distribution (SED) from the ultraviolet to far-ultraviolet or even X-ray wavebands. Therefore, the emission lines carry the information of the part SED of AGNs \citep[e.g.,][]{Boroson1992}. \cite{Korista2004} explored the effects of a global change in the ionizing continuum level on the behavior of the stronger optical hydrogen and helium broad emission lines based on a distributed cloud photoionization model of a BLR in the LOC scenario \citep{Baldwin1995}. They found that the differing responsivities$({\rm dlog}L_{\rm line}/{\rm dlog}L_{\rm cont})$ among the emission lines can explain several observed phenomena, including the anti-correlation of Balmer decrement with continuum state within an individual AGN. However, they neither considered an evolution of the SED with a continuum state (which can further change Q(H)), nor SED differences across the populations of AGN with different masses and $L/L_{\rm Edd}$. These considerations are central to the present study. \cite{Ferland2020} also investigated the relationship between the hydrogen and helium broad emission lines and AGN SEDs, and proposed that systematic changes in the SEDs with  Eddington ratio alone cannot fully explain the hydrogen Balmer to He {\footnotesize II} emission line flux ratios and that the integrated BLR covering factor may be decreasing systematically with increasing Eddington ratio. \cite{Guo2020} studied the variability of Mg {\footnotesize II}, H$\alpha$, and H$\beta$ lines in changing-look quasars and proposed that the dramatic changes of the broad lines are consistent with photoionization responses to extreme continuum variability.

In recent years, accumulating observational evidence supports the scenario that CLAGNs undergo strong changes in continuum state \citep[e.g.,][]{Noda2018}, as well as potentially accretion mode transitions during the changing look phase \cite[e.g.,][]{2019ApJ...874....8M,Guolo2021,Lyu2021,Liu2022}. In this work, we explore the possible influence of the evolution of ionizing SED on the Balmer decrement in CLAGNs and LLAGNs, which can shed light on the physical reason for the higher Balmer decrement and the possible physics of the disappearance/re-appearance of broad emission lines. In Section \ref{sec:sample}, we present our sample. The model and results are presented in Sections \ref{sec:model} and \ref{sec:result}, respectively. Conclusion and Discussion are shown in Section \ref{sec:discussion}. In this work we use the standard cosmological model ($H_0={\rm 70 \, km \, s^{-1}\, Mpc^{-1}}, \Omega_\Lambda=0.7,\Omega_{\rm M}=0.3$).

\section{Sample}\label{sec:sample}
There are many reported AGN samples for exploring the Balmer decrements in the literature, and most of them focus on the bright blue AGNs. According to the purpose of our work, we need an AGN sample with a wider distribution of Eddington ratios. In this work, our sample consists of normal AGNs and changing-look AGNs with measurements of both $\rm H\alpha$ and $\rm H\beta$ intensities.

We select AGNs from \cite{LaMura2007} and \cite{Jaffarian2020}, which include AGNs with a wider distribution of Eddington ratios (e.g., type 1-1.9 or even some true type 2 sources with weak absorption). There are 90 AGNs in \cite{LaMura2007}, which have a black hole (BH) mass in the range of $10^{5-9}M_{\odot}$ and bolometric Eddington ratio ($\lambda_{\rm Edd}=L_{\rm bol}/L_{\rm Edd}$) in range of 0.01-1. We also select 33 AGNs with bolometric Eddington ratio $0.001\le \lambda_{\rm Edd}\le 1$ and low X-ray column density (i.e. $N_{\rm H} \le 10^{21} {\rm cm^{-2}}$) from \cite{Jaffarian2020}, which imply no prominent intrinsic reddening and extinction. The bolometric Eddington ratio is estimated from the optical \citep[e.g., $L_{\rm bol} \sim 9L_{\rm 5100}$,][]{Kaspi2000,Elvis1994} or X-ray observations with the empirical bolometric corrections (see Table \ref{table1} and references).  

For CLAGNs, we include 6 sources with multiple optical spectral observations (NGC 5548, NGC 4151, 3C 390.3, NGC 3516, Mrk 1018, and 1ES 1927+654) and 16 sources with only two observations from \cite{Jin2022}. We note that only the emission-line intensities at a high state (or turn-on state) with evident broad lines are adopted in our work because the emission lines are weak or absent in a low state (or turn-off state). The multiple observational data for NGC 5548, NGC 4151, and 3C 390.3 are selected from \cite{Raki2017}. For Mrk 1018, we adopt quasi-simultaneous observation of $\rm H\alpha$ and $\rm H\beta$ lines from \cite{Cohen1986} and \cite{McElroy2016}. The emission line data of NGC 3516 are selected from \cite{Shapovalova2019}, where we exclude the data from 2014 to 2018 due to the $\rm H\beta$ line being too weak and the measurements on the broad component are not accurate. The data of 1ES 1927+654 is selected from \cite{Ruancun2022}. For CLAGNs with only two spectra in \cite{Jin2022}, we exclude the objects observed by XinLong Telescope for the bad spectral quality and several sources with very large differences of velocity even in active state (i.e., $\rm FWHM(H\beta)/FWHM(H\alpha)\ge1.5$ or $\rm FWHM(H\alpha)/FWHM(H\beta)\ge 1.5 $). The source ZTF18aajupnt is considered to be a candidate for tidal disruption event \citep{Jin2022}, which is also removed from our sample. The BH mass ${\rm log}M_{\rm BH}$=8.61, 7.72, 7.81, 7.28, 7.70, and 7.90 are adopted for the CLAGN of 3C 390.3 \citep{Afanasiev2019}, NGC 4151 \citep{Du2015}, NGC 5548 \citep{Afanasiev2019}, 1ES 1927+654 \citep{Trakhtenbrot2019}, NGC 3516 \citep{Vasudevan2010} and Mrk 1018 \citep{Noda2018}, respectively. It should be noted that Mrk 590 is a well-known CLAGN, which is not included in our AGN sample because we only find one simultaneous flux of $\rm H\alpha$ and $\rm H\beta$ lines that are reported in the literature.

\begin{deluxetable*}{lcclccc}
\centering
\tabcolsep=0.6cm
\tablecaption{Objects with low absorption of $N_{\rm H}<10^{21}\rm cm^{-2}$ selected from \cite{Jaffarian2020}. \label{table1}}
\tablehead{
\colhead{Objects} & \colhead{z} & \colhead{H$\alpha$/H$\beta$} & \colhead{${\rm log}\lambda_{\rm Edd}$} & \colhead{log$M_{\rm BH} (M_\odot)$} & \colhead{log$N_{\rm H}$} & \colhead{Reference}
}
\colnumbers
\startdata
Mrk 110 & 0.035 & 4.22 & -1.89 & 8.32 & 20.17 & 1 \\
I Zw 1  & 0.061 & 4.66 & -1.13 & 7.46 & 20.81 & 1 \\
Mrk 335 & 0.026 & 2.60 & -0.92 & 7.49 & 20.55 & 1 \\
Mrk 79  & 0.022 & 5.53 & -1.00  & 7.45 & 20.97 & 1 \\
Mrk 1044 & 0.016 & 2.40 & -0.49 & 6.45 & 20.62 & 2 \\
Mrk 1310 & 0.020 & 3.97 & -1.48 & 6.62 & 20.48 & 2 \\
Mrk 142 & 0.045 & 2.90 & -0.17  & 6.59 & 20.40 & 2 \\
Mrk 205 & 0.071 & 4.28 & -1.52  & 8.32 & 20.35 & 14 \\
Mrk 304 & 0.066 & 2.86 & -0.80  & 8.04 & 20.16 & 1 \\
Mrk 40  & 0.021 & 2.65 & -1.19* & 7.09 & 20.36 & 7,16 \\
Mrk 42  & 0.024 & 3.63 & -0.04*  & 6.00 & 19.90 & 7,8 \\
Mrk 474 & 0.037 & 3.09 & -1.04* & 7.41 & 20.49 & 7,16 \\
Mrk 493 & 0.031 & 2.71 & -0.17  & 6.14 & 20.47 & 2 \\
Mrk 876 & 0.121 & 5.20 & -0.80  & 8.36 & 19.63 & 1 \\
NGC 985 & 0.043 & 4.67 & -1.24 & 8.36 & 20.77 & 4 \\
PG 1001+054 & 0.160 & 3.20 & -0.30 & 7.74 & 19.37 &  6 \\
PG 1244+026 & 0.048 & 2.95 & -0.10 & 7.21 & 19.49 & 9 \\
PG 1448+273 & 0.064 & 3.40 & -0.12 & 7.29 & 19.64 & 10 \\
Fairall 9 & 0.046 & 2.59 & -1.26 & 8.09 & 20.30 & 2 \\
Mrk 50  & 0.024 & 3.30  & -0.10* & 6.18 & 19.90 & 7,16 \\
UGC 6728 & 0.007 & 7.00 & -1.04 & 5.32 & 19.74 & 11 \\
Mrk 509 & 0.034 & 2.69 & -1.46 &  8.56 & 20.81 & 4 \\
Mrk 817 & 0.031 & 4.00 & -1.11 &  7.65 & 20.06 & 1 \\
NGC 1566 & 0.005 & 4.18 & -1.57* & 6.65 & 19.73 & 7,16 \\
Mrk 1218 & 0.029 & 9.04 & -2.96 & 8.75 & 20.65 & 5,16 \\
Mrk 841 & 0.036 & 4.80 & -1.62 & 8.76 & 20.20 & 1 \\
Mrk 609 & 0.034 & 4.95 & -1.48 & 7.80 & 20.77 & 12 \\
NGC 3660 & 0.012 & 7.50 & -2.26 & 6.85 & 20.26 & 15 \\
NGC 5273 & 0.004 & 4.24 & -2.74 & 7.14 & 20.95 & 2 \\
NGC 5929 & 0.009 & 5.13 & -0.89 & 7.25 & 20.71 & 3 \\
NGC 7590 & 0.005 & 3.65 & -1.40 & 6.51 & 20.96 & 3 \\
Mrk 266SW & 0.028 & 6.88 & -2.04 & 8.30 & 20.61 & 13 \\
Mrk 590 & 0.026 & 4.19 & -1.19 & 7.55 & 20.54 & 2 \\
\enddata
\tablecomments{References: (1) \cite{Afanasiev2019}; (2) \cite{Du2015}; (3) \cite{Bian2007}; (4) \cite{Vasudevan2010}; (5) \cite{Singh2011}; (6) \cite{Wangchaojun2022}; (7) \cite{Rudy1984}; (8) \cite{Wangt2001}; (9) \cite{Jin2013}; (10) \cite{Laurenti2021}; (11) \cite{Garc2019}; (12) \cite{Lyu2022}; (13) \cite{Iwasawa2020}; (14) \cite{Laha2019}; (15) \cite{Rivers2016}; (16) The BH mass is calculated from the $M-\sigma_{*}$ relation \cite{Ferrarese2000} where the dispersion $\sigma_*$ is adopted from the HyperLeda database \url{http://leda.univ-lyon1.fr/}. The stars in column (4) indicate that $L_{5100}$, from which $L_{\rm bol}$ is estimated, was calculated from $L_{\rm H\alpha}$ using $L_{\rm H\alpha}-L_{\rm 5100}$ as proposed by \cite{Greene2005}.}
\end{deluxetable*}

\section{Photoionization Model}\label{sec:model}

To explore the properties of broad emission lines, we simulate a grid of photoionization models based on Ferland's spectral synthesis code Cloudy \citep[version 17.03,][]{Ferland2017} and consider a wide variation of the incident continuum SEDs associated with ranges in the mass of SMBH and the Eddington ratio. More details will be presented below.

\begin{figure}[ht!]
\includegraphics[scale=0.65]{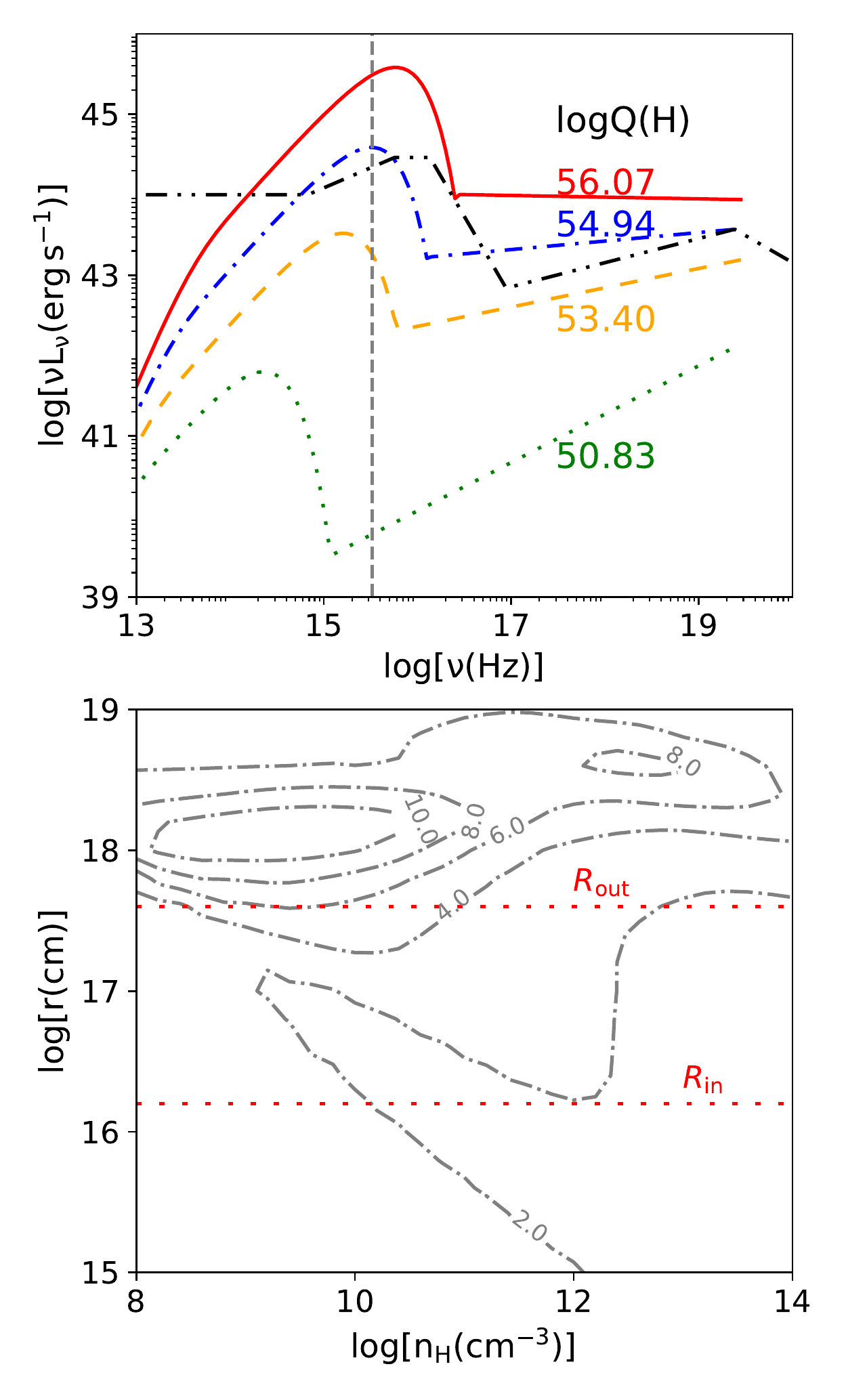}
\caption{The upper panel shows the incident SED for a typical BH mass of $M_{\rm BH}=10^8 M_\odot $, which includes a blackbody component and a power-law component with cutoff at ${\rm log} \nu =19.5$. The red solid line and the blue dot-dashed line represent the SED with no disk truncation with dimensionless accretion rate $\dot{m}=$ 1 and 0.1, respectively, while the orange dashed line and the red dotted line represent the case for the truncated disk with $ R_{\rm tr}=12 R_{\rm g}$ at $\dot{m}=0.01$ and $120 R_{\rm g}$ at $\dot{m}=0.001$, respectively. Typical quasar SED that is widely adopted in the photoionization model is plotted with black double dots dashed line \citep{Mathews1987}. The vertical grey dashed line represents the position of 1 Ryd. The number of hydrogen ionizing photons emitted by the source per second, Q(H), is presented for corresponding SEDs. The lower panel shows the contour map of $\rm H\alpha/H\beta$, as predicted by the photoionization model clouds in the grid, where the SED corresponds to the blue dot-dashed line in the upper panel. The red dotted lines show the outer and inner boundary of BLR that correspond to the case of $\dot{m}=0.1$ and ${\rm log Q(H)=54.94}$, the blue dot-dashed line in the top panel.
\label{fig1}}
\end{figure}

\subsection{Incident continuum}
It is found that the blackbody component and power-law component are strongly evolved with the change of the Eddington ratios in both AGNs and BH X-ray binaries (XRBs). Apart from the energy dissipation in the SSD, a small fraction of the gravitational energy will be released into the corona, which is needed to explain the power-law component as found in both high-state XRBs and luminous AGNs. The SSD contribution will become weak and the power-law component will become dominant in low-state XRBs and LLAGNs, where the cold SSD has possibly transitioned into the hot ADAF when the accretion rate is lower than a critical value \citep[e.g., $\dot{m}_{\rm c}=\dot{M}/\dot{M}_{\rm Edd}\sim 2\%$,][]{Yuan2014}. In this work, we simply adopt that the transition radius follows $R_{\rm tr} = 6 (\dot{m}/\dot{m}_c)^{-1} R_{\rm g}$ when the accretion rate is lower than the critical value \citep[$R_{\rm g}=G M/c^2$ is the gravitational radius,][]{Taam2012}. The blackbody emission is calculated based on either the full SSD or the outer SSD in the truncated disk. The physical formation of the corona/ADAF is still unclear, and we calculate the power-law component from two observed empirical correlations: 1) the bolometric correction on the 2-10 keV luminosity $k_{\rm bol}{\rm (2-10keV)}=7 \times [L({\rm 2-10keV})/10^{42} {\rm erg \, cm^{-2}}]^{0.3}$ by \cite{Netzer2019}; 2) relation between X-ray photon index and bolometric luminosity $\Gamma _{2-10\rm{keV}}-L_{\rm bol}$ \citep{Qiao2018}. 

In the upper panel of Figure \ref{fig1}, we present examples of incident SED with a blackbody component and a power-law component with $\dot{m}=0.001-1$ for $M_{\rm BH}=10^8 M_\odot$. For comparison, the typical SED widely adopted in the photoionization model in the literature is also presented.

\subsection{Photoionization models and line emission}

Locally Optimally Emitting Cloud (LOC) scenario as proposed by \cite{Baldwin1995} can successfully reproduce the emission-line strengths and line ratios in both BLR and NLR of different types of AGNs \citep[e.g.,][]{Ferland2003,Ferguson1997,Korista2000}. In this scenario, only a relatively narrow range of parameters (e.g., gas density and incident hydrogen ionizing photon flux and their ratio that is the ionizing parameter) will result in high continuum reprocessing efficiencies for a particular emission line. Thus a broad range in such parameters is then indicated to explain the strengths of the wide range of commonly found broad emission lines in AGN. The integrated emission line luminosity can be approximated as \citep[see also][]{Bottorff2002} 
\begin{equation} \label{equ1}
    L_{\rm line} \propto \int{\int{r^2 F(r,n) f(r) g(n) dn dr}},
\end{equation}
where $F(r,n)$ is the emission line surface flux of a cloud at distance $r$ from the central ionizing source with constant density $n({\rm H})=n$, for a choice of a total column density $N({\rm H})$. The functions $f(r)$ and $g(n)$ are weighting functions, which are normally chosen as power laws in each variable. The function $f(r)\sim r^{\Gamma}$ represents the radial differential covering fraction ${\rm d}C(r)/{\rm d}r$ of the line-emitting clouds, while function $g(n)\sim n^{\beta}$ describes the weighting of clouds of differing gas densities to the total solid area covered by clouds at distance $r$. We calculate the line luminosity (i.e., $L_{\rm line}$) by integrating over a logarithmic grid of photoionization models in parameter space of $(r,n)$, with proper outer boundaries. In this work, we adopt the typical values of $\Gamma = -1, \beta = -1$, as proposed in \cite{Baldwin1995}, \cite{Korista2000} and \cite{Nagao2006}.

Based on the reverberation mapping observations, the typical BLR size is correlated with the luminosity \citep{Blandford1982,Peterson1993,Peterson2004,Bentz2009}. However, the inner and outer boundaries of BLR are still unclear, even though the torus size is sometimes adopted as the outer BLR boundary \citep[$R_{\rm torus}\sim 5 R_{\rm BLR}$,][]{Koshida2014}. \cite{Naddaf2020} found that the ratio $R_{\rm out}/R_{\rm in}\sim20-60$ for $\dot{m}=0.01$ to 1 by assuming that the BLR clouds are formed from the Failed Radiatively Accelerated Dusty Outflow (FRADO). In this work, we simply set $R_{\rm out}/R_{\rm in}=40$, and the other boundary conditions will be discussed. The volume density of BLR clouds should be larger than $\sim 10^9 {\rm cm}^{-3}$, due to no observed broad forbidden lines, e.g. [O {\footnotesize III}] $\lambda 4363$. \cite{Schnorr2016} gave the BLR hydrogen density $\sim 10^{11} \rm{cm}^{-3}$ by studying some nearby Seyfert galaxies\citep[see also,][]{Ferland1992,Shields1993}. \cite{Korista2000} proposed an upper limit of the BLR density $\sim 10^{12} {\rm cm}^{-3}$, since clouds of higher gas densities transition to thermal continuum emitters rather than sources of emission lines \citep[see also][]{Guo2020}. In this work, we set the BLR hydrogen density to vary from $10^{8} {\rm cm^{-3}}$ to $10^{12} {\rm cm^{-3}}$ with grid steps of 0.2 dex in simulations. Finally, we adopt a fixed value in the cloud total hydrogen column density $N_{\rm H}=10^{23} {\rm cm^{-2}}$, as representative of such within the BLRs of AGN \citep{Dumont1998}, and assume solar gas elemental abundances $Z=Z_{\odot}$. To calculate the line luminosity, the typical cloud covering factor 0.5 is adopted in this work \citep{Korista2000,Guo2020}, which may decrease as the accretion rate increases \citep{Ferland2020}. We note that while the integrated covering factor affects the predicted emission line luminosities, it does not affect the predicted emission line luminosity ratio.

In the lower panel of Figure \ref{fig1}, we present an example of a contour map of $\rm H\alpha/H\beta$ in the density-distance plane with a dimensionless accretion rate $\dot{m}=0.1$. The contours are those of the flux ratio $F({\rm H\alpha})/F({\rm H\beta})$ predicted by the photoionization model clouds in the grid. The larger radius will lead to a smaller hydrogen-ionizing photon flux ${\rm \Phi(H)=Q(H)/4\pi} r^2$ and a steeper Balmer decrement \citep[see also][]{Korista2004}. With the above typical model parameters, we present several typical line luminosities in Table \ref{table2}, and compare the theoretical prediction of H$\alpha$ and H$\beta$ lines with the observations (e.g., the observed relation of $L_{5100}-L_{\rm H\alpha}$ and $L_{5100}-L_{\rm H\beta}$), where the line luminosities are roughly consistent with those from observations. For the line ratios, our model predicts $\rm Ly\alpha/H\beta \sim 20$, Mg {\footnotesize II} $\lambda2800/\rm H\beta \sim 1.5$,  C {\footnotesize IV} $\lambda1549/\rm H\beta \sim 4$ and He {\footnotesize II} $\,\lambda1640/\rm H\beta \sim 0.5$ for above typical parameters with $\dot{m}\sim0.1$. This is roughly consistent with the observations of three Seyferts with $L_{\rm bol}/L_{\rm Edd}\sim 0.1$, where all above lines are observed and the line ratios are $\rm Ly\alpha/H\beta\gtrsim 8-18$, Mg {\footnotesize II} $\lambda2800/\rm H\beta\sim 1-2$, C {\footnotesize IV} $\lambda 1549/\rm H\beta\sim 2-8$  and He {\footnotesize II} $\lambda 1640/\rm H\beta\sim 0.4-0.7$ respectively \citep{Marziani2010}. Therefore, our model can roughly reproduce the observational line luminosities and line ratios with the typical model parameters.

\begin{figure*}[ht!]
\includegraphics[scale=0.85]{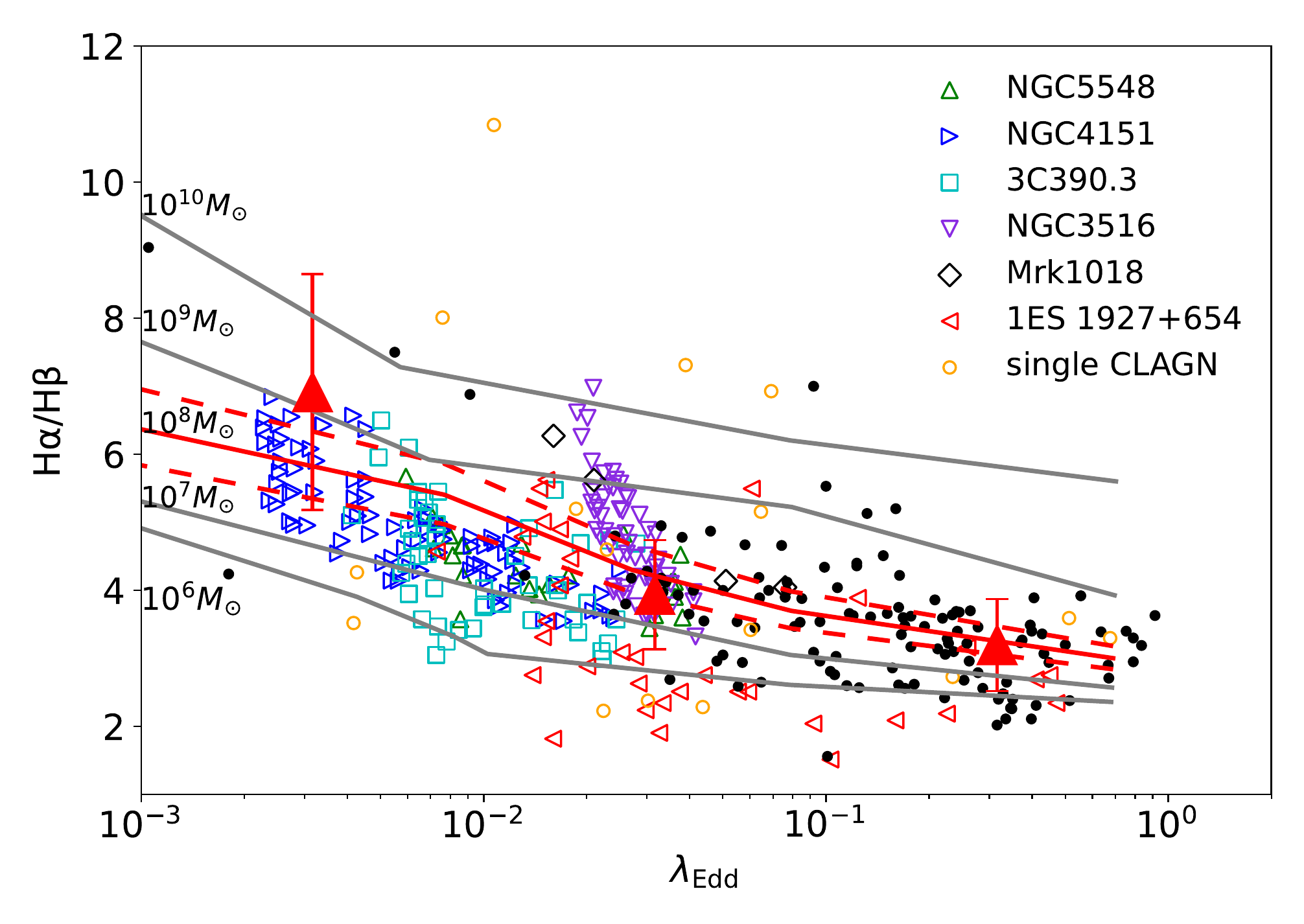}
\caption{The correlation of $\rm H\alpha/H\beta-\lambda_{\rm{Edd}}$. The black dots represent the AGN sources selected from \cite{LaMura2007} and \cite{Jaffarian2020}. The open symbols represent the CLAGNs with multiple observations, and the orange open circles represent the active state of some CLAGNs selected from \cite{Jin2022}. The red solid triangles with error bars represent the average ratios of $\rm H\alpha/H\beta$ for the sources in three bins ($10^{-3}<\lambda_{\rm Edd}<10^{-2},10^{-2}<\lambda_{\rm Edd}<10^{-1},10^{-1}<\lambda_{\rm Edd}<1$). The solid lines from top to bottom represent the theoretical results for $10^{10} M_{\odot}, 10^9M_{\odot}, 10^8 M_{\odot}, 10^7 M_{\odot}$ and $10^6 M_{\odot}$, respectively for $R_{\rm out}/R_{\rm in}=40$. For comparison, we also represent two cases with $R_{\rm out}/R_{\rm in}=60$ (upper dashed line) and 20 (lower dashed line) for $M_{\rm BH}=10^8M_{\odot}$.
\label{fig2}}
\end{figure*}

\section{Result}\label{sec:result}

In Figure \ref{fig2}, we present the relation between $\rm H\alpha/H\beta$ and $\lambda_{\rm{Edd}}$ for our sample. It can be found that the sources with low Eddington ratios have steeper Balmer decrement, where $\rm{H\alpha/H\beta}=4-10$ when the Eddington ratio is less than a few percent, while the average value of $\rm{H\alpha/H\beta}$ is 3.24 for bright AGNs with $\lambda_{\rm{Edd}}\ge 0.1$. The Spearman correlation coefficient is $r=-0.51$ with $p=1.32 \times 10^{-9}$ for only AGN sample, and $r=-0.65$ with $p<1\times 10^{-20}$ for all sources including CLAGNs. This negative trend is also found in a single CLAGN, but its slope may be different for different sources. The average Balmer decrement for 16 CLAGNs in turn-on state from \cite{Jin2022} is $\rm{H\alpha/H\beta}=4.73$, which is also statistically higher than that of normal bright AGNs. Except for the evident deviation of $\rm H\alpha/H\beta$ for sources with $\lambda_{\rm{Edd}}<$1-2\%, the slope of the $\rm H\alpha/H\beta-\lambda_{\rm{Edd}}$ correlation also become steeper in CLAGNs of 3C 390.3 and 1ES 1927+654 at lower Eddington ratios. Therefore, the change of the slope in $\rm H\alpha/H\beta - \lambda_{\rm{Edd}}$ correlation is found in both single CLAGNs and AGN samples.

In Figure \ref{fig2}, we also present the theoretical prediction for $\rm H\alpha/H\beta-\lambda_{\rm{Edd}}$, where the solid lines represent the $M_{\rm{BH}}=10^6, 10^7, 10^8, 10^9, 10^{10} M_{\odot}$ (from bottom to top), respectively. For given BH mass, the correlation becomes steeper when the standard disk to hot ADAF at low Eddington ratios, which is roughly consistent with the observations. For a given Eddington ratio, we find that more massive BHs predict higher values of $\rm{ H\alpha/H\beta}$, where $\rm{H\alpha/H\beta}=6-8$ for $M_{\rm BH}=10^{10} M_{\odot}$ and $\rm{H\alpha/H\beta}=2.6-3.3$ for $M_{\rm BH}=10^{6} M_{\odot}$ in the case of $\lambda_{\rm{Edd}}=0.01-1$. We find that the BLR boundary also slightly affects the results (see the dashed line in Figure \ref{fig2}), where the larger value $R_{\rm out}/R_{\rm in}$ predicts a slightly steeper Balmer decrement. The average logarithmic mass of our whole sample is 7.2 with a standard deviation of 0.7, and most sources are roughly in the range of $M_{\rm{BH}}=10^6 - 10^8 M_{\odot}$. It should be noted that the scatter in observational relation is still large (e.g., $\rm H\alpha/H\beta\sim 5$ for some bright AGNs with $\lambda_{\rm{Edd}}\sim 0.1$), where better observational sample with wider distribution of BH masses is wished to further explore this relation.

\begin{figure*}[ht!]
\includegraphics[scale=0.75]{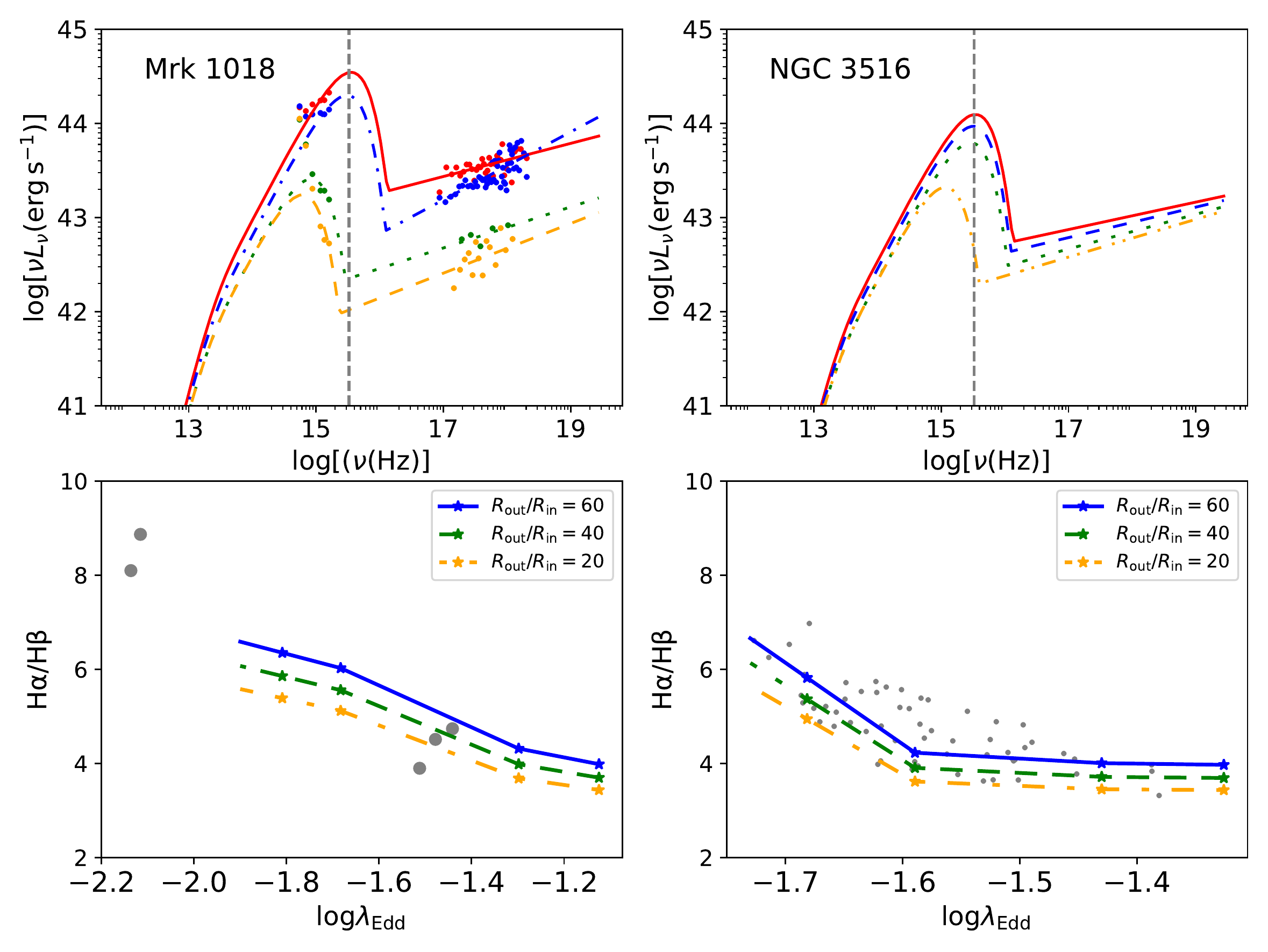}
\caption{The top-left panel presents the SEDs of Mrk 1018 based on modeling the multi-wavelength quasi-simultaneous observations and the red solid, blue dot-dashed, green dotted, orange dashed lines represent the $(\dot{m},R_{\rm tr})=$ $(0.11,6R_{\rm g})$, $(0.07,6R_{\rm g})$, $(0.03,70R_{\rm g})$ and $(0.03,110R_{\rm g})$, which correspond to ${\rm \log Q(H)}=54.92, 54.61, 51.67, 50.27$, respectively. The grey dashed line represents the position of 1 Ryd. The Balmer decrement calculated from the photoionization model based on the above SEDs and comparison with observations is shown in the bottom-left panel. The grey dots are the observation data, and the green stars are calculated by simulation. The right two panels represent the result for NGC 3516. In the right upper panel, the red solid, blue dashed, green dotted, and orange double dots dashed lines represent the $(\dot{m}, R_{\rm tr})=(0.07,6R_{\rm g}),(0.05,6R_{\rm g}),(0.03,6R_{\rm g}),(0.03,50R_{\rm g})$, which correspond to ${\rm \log Q(H)}=54.47, 54.31, 54.08, 52.51$, respectively. In the lower panels, the blue solid, green dashed, and orange dot-dashed lines are calculated by varying the BLR boundaries $R_{\rm out}/R_{\rm in}= 60, 40, 20$, respectively.
\label{fig3}}
\end{figure*}

We further explore the correlation of ${\rm H\alpha/H\beta}-\lambda_{\rm Edd}$ in two CLAGNs (Mrk 1018 and NGC 3516). Mrk 1018 shows strong evolution of SEDs in four quasi-simultaneous observations from optical to X-ray \citep[e.g.,][and references therein]{Noda2018,Lyu2021}. To build the SED for each observation, we fit the optical spectrum with SSD or truncated SSD model and fit the X-ray spectrum with a power-law component, where we simply assume the cutoff at ${\rm log} \nu= 19.5$. We only fit four optical data points in optical-UV bands because the contribution from a galaxy is not important in these bands. For two bright states, the SSD can roughly reproduce the optical spectra, while two low-state observations need the truncated SSD with a transition radius of $70 R_{\rm g}$ and $110 R_{\rm g}$, respectively (see the top-left panel of Figure \ref{fig3}). We calculate the ratio of $\rm H\alpha/H\beta$ based on the fitted SEDs from observations, where the different BLR boundaries ($R_{\rm out}/R_{\rm in}$) are also considered(see the left-bottom panel in Figure \ref{fig3}). For NGC 3516, we don't have simultaneous optical to X-ray data and we build the SED based on the optical data together with the empirical correlations $\Gamma _{2-10\rm{keV}}-L_{\rm bol}$ relation\citep{Qiao2018} and the theoretical relation $k_{\rm bol}{\rm (2-10keV)}-L{\rm (2-10keV)}$ by \cite{Netzer2019}, where $k_{\rm bol}{\rm (2-10keV)}$ is the bolometric correction factor for X-ray luminosity in 2-10keV.  It should be noted that ${\rm H\alpha/H\beta}-\lambda_{\rm Edd}$ relation in NGC 3516 is a little bit steeper compared with the other CLAGNs (see Figure \ref{fig2}), where the parameters of $(\dot{m},R_{\rm tr})=(0.07,6R_{\rm g}), (0.05,6R_{\rm g}), (0.03,6R_{\rm g}), (0.03,50R_{\rm g})$ are adopted to reproduce the observed correlation.

\section{Conclusion and Discussion}\label{sec:discussion}
After including more AGNs with low-Eddington ratios, we find that the intensity ratio of $\rm H\alpha$ to $\rm H\beta$ is strongly anti-correlated to the Eddington ratios in both AGN sample and in individual CLAGNs. We model the Balmer decrement based on the photoionization model using the Cloudy code, where the incident AGN SEDs are assumed to be evolved with the accretion rates. Considering the possible disk evolution, we find that the observed anti-correlation can be reproduced by the theoretical model. We propose that the SEDs with lower accretion rates can also lead to a higher ratio of $\rm H\alpha/H\beta$, where the steeper Balmer decrement may be not fully caused by the internal reddening since some LLAGNs and CLAGNs have negligible absorption.

The Balmer lines are the most well-known and strong emission lines in the optical spectrum,  which are sometimes used to shed light on the amount of dust extinction due to its insensitivity to the gas conditions \citep[e.g., temperature and density,][]{Osterbrock1989}. 
However, the Balmer decrement of broad lines cover a wide range from one sample to another (e.g., $\rm H\alpha/H\beta=3.06\pm0.22$ based on 446 blue Seyfert 1 from the SDSS, \citealp{Dong2008}; $3.2\pm0.6$ for 554 bright AGNs in SDSS DR7, \citealp{Lu2019}; $3.45\pm0.65$ for 90 nearby Seyfert 1, \citealp{LaMura2007}). \cite{LaMura2007} and \cite{Lu2019} found that the Balmer decrement may be anti-correlated with the accretion rate or Eddington ratio, where they attributed to the possible stronger absorption with accretion rate decreases. After including more sources with low-Eddington ratios and some CLAGNs, we find this anti-correlation becomes stronger, where the Spearman correlation coefficient $r=-0.65$ with $p< 10^{-20}$. It should be noted that the steep Balmer decrement in some sources with low Eddington ratios might be not completely caused by the absorption since that many types I sources have $\rm H\alpha/H\beta=5-10$ but with little absorption of $N_{\rm H}<10^{21}\rm cm^{-2}$ \citep[e.g.,][]{Jaffarian2020,Crenshaw1988,Osterbrock1976}. Furthermore, the anti-correlation of ${\rm H\alpha/H\beta}-\lambda_{\rm Edd}$ is found in most CLAGNs with a strong variation of optical emission lines, where the decrease of broad $\rm H\beta$ line is normally faster than $\rm H\alpha$ line (\citealp[e.g., Mrk 1018,][]{Kim2018} and \citealp[1ES 1927+654,][]{Ruancun2022}). More importantly, the absorption is always weak in some CLAGNs \citep[e.g. Mrk 1018,][]{LaMassa2017} during the changing look stage. \cite{Mehdipour2022} explored the possible physical reason for the changing look phenomenon in NGC 3516 based on optical to X-ray observations, where they found its low-flux spectrum variability does not require any new or variable obscuration. Therefore, the variation of Balmer decrement cannot be fully attributed to the variation of absorption even though we cannot exclude it in all sources.

The properties of broad emission lines in AGNs are also affected by the continuum state, where the changes in AGN SEDs will affect the flux of photons for ionization. The evolution of the broad emission lines in a single CLAGN also supports this scenario. In CLAGNs, the $\rm H\beta$ lines normally become weak faster than the $\rm H\alpha$ lines during the decay of CLAGNs. This is similar to the case that the broad $\rm H\alpha$ lines are observed in some LLAGNs and the broad $\rm H\beta$ lines are weak or disappear. The UV ionization photons will quickly decrease with the decrease in accretion rate, where the UV emission will further decrease if the inner SSD is truncated to ADAF when the accretion rate is lower than a critical value. Based on the putative evolution of nuclear SEDs with the variation of the accretion rate, our model naturally predicts the anti-correlation of $\rm H\alpha/H\beta-\lambda_{\rm Edd}$, where the $\rm H\beta$ line decreases faster due to the fast decay of UV photons in the case of low-accretion rate. The anti-correlation of $\rm H\alpha/H\beta-\lambda_{\rm Edd}$ is most possibly triggered by the optical depth effects that have been mentioned in earlier works \citep[e.g.,][]{Netzer1975,Rees1989,Korista2004}. The optical depths of the Balmer emission lines within the BLR of an AGN are generally proportional to the flux of hydrogen ionizing photons, and will therefore correlate with changes in the AGN's ionizing photon luminosity Q(H). At a given luminosity of the central source, the optical depth of $\rm H\alpha$ is larger than that of $\rm H\beta$. Thus, an increase in the Q(H) will lead to a relatively smaller increase in $\rm H\alpha$ photons escaping the clouds than $\rm H\beta$ photons. That is, the responsivity $({\rm dlog}L_{\rm line}/{\rm dlog}L_{\rm cont})$ of $\rm H\alpha$ is expected to be smaller than that of $\rm H\beta$ (which is smaller than that in $\rm H\gamma$), and this results in a flattening of the Balmer decrement in high continuum states and steepening in low continuum states. The anti-correlation shown in Figure \ref{fig2} or Figure \ref{fig3} can also be understood by referring to the lower panel of Figure \ref{fig1}. How the gas emits the Balmer lines depend mostly on the flux of hydrogen ionizing photons $\rm \Phi(H)$ striking on the cloud. So if the source changes from a higher continuum state to a lower continuum state, the contours in this panel will slide down along the distance-axis owing to ${\rm \Phi(H)\propto Q(H)}/r^2$. But the $R_{\rm BLR}$ relying on $L_{5100}$ decrease slower than the flux of hydrogen ionizing photons which have higher energy. This leads to the contours sliding down relative to the BLR boundaries (i.e., the red dashed lines) in the panel. Finally, the whole BLR will emit a steeper Balmer decrement than the case in which the source is in the high state.

\cite{Korista2004} found that their BLR model predicts the broad $\rm H\alpha/H\beta$ flux ratio to change from 4.9 in a low continuum state to 3.7 in a high continuum state, for the far-UV luminosity of NGC 5548 changes a factor of 8 for over 13 years period. In this, they assumed a constant SED, such that Q(H) scales 1:1 with the observed far-UV luminosity (although they discussed possible impacts on the emission line responsivities for a variable SED). Had they explored quantitatively changes in the SED with continuum state, such that Q(H) is further enhanced in high continuum states and further reduced in low continuum states, they would have found a greater range in the flux ratio of $\rm H\alpha/H\beta$ -- likely even closer to the range observed in NGC 5548 (see Figure \ref{fig2}). It should be noted that the physical mechanism for the SSD transition is still unclear, and we adopt a simple relation of $R_{\rm tr}-\dot{m}$ in calculating the blackbody emission. For the X-ray spectrum, we adopt the empirical correlation of $\Gamma-L_{\rm bol}$ and theoretically calculated relation $k_{\rm bol}{\rm (2-10keV)}-L{\rm (2-10keV)}$, which will not affect our main conclusion. For a given mass black hole, the number of ionizing photons emitted per second Q(H) is strongly correlated with the Eddington ratio, where UV photons become weaker or roughly disappear as a decrease in accretion rate (in particular for the case of a truncated disk). The parameter Q(H) decreases much faster if the inner SSD transition to ADAF since UV photons dominantly originate from the cold SSD (e.g., see top panel of Figure \ref{fig1}). The correlation of $\rm H\alpha/H\beta-\lambda_{\rm Edd}$ becomes steeper for $\lambda_{\rm Edd}<$1\%, which is mainly caused by the faster decrease of $\rm Q(H)$ in the ADAF case. NGC 3516 follows a steeper correlation of $\rm H\alpha/H\beta-\lambda_{\rm Edd}$ with $\log\lambda_{\rm Edd}<-1.5$, which may be triggered by a little bit higher critical accretion rate for disk transition (see Figure \ref{fig3}). Furthermore, the relation of $R_{\rm tr}-\dot{m}$ may be different from source to source, which may be one the reason why some CLAGNs follow their own $\rm H\alpha/H\beta-\lambda_{\rm Edd}$ track.

We note that our model also predicts that the AGNs with more massive BHs will have larger $\rm H\alpha/H\beta$ (see Figure \ref{fig2}), where the $\rm H\alpha/H\beta$ is expected to be 5-6 for a black hole with $M_{\rm BH}= 10^{10}M_{\odot}$, even if they have high accretion rates (e.g., $\ge 0.1 \dot{M}_{\rm Edd}$). This is caused by the peak energy of the SEDs from SSD moving into sub-ionizing UV band, similar to the cases of low accretion rates or truncated disks for lower masses. Further observational tests with more massive BHs will be the subject of future works.

The LOC model is adopted to explore the properties of emission lines in this work, where the line emission originates from the combination of all clouds but is dominated by those with the highest efficiency of reprocessing the incident ionizing continuum \citep{Baldwin1995}. The LOC models provide good fits to the measured line fluxes. Similar to the traditional LOC model, we simply set $\Gamma=-1$ and $\beta=-1$ for the distribution functions of clouds. \cite{Korista2000,Nagao2006} found that $\Gamma$ value may be slightly smaller than the typical value of -1 based on the fitting of emission lines (e.g.,$\Gamma\sim -1.4$ to $-1$). We find that the Balmer line luminosities will increase about 0.6 times and the line ratio $\rm H\alpha/H\beta$ will increase about 0.4 when $\Gamma$ varies from -1.4 to -1, which is mainly caused by the weight biasing to the inner part of clouds within the BLR. It should be noted that a slight change of $\Gamma$ and $\beta$ will not change our conclusion on the anti-correlation of $\rm H\alpha/H\beta-\lambda_{\rm Edd}$ correlation. Apart from the parameters of $\beta$ and $\Gamma$, the line luminosities and line ratios are also correlated with the inner/outer BLR radius in the integration. In Figures \ref{fig2} and \ref{fig3}, we can find that the larger value of $R_{\rm out}/R_{\rm in}$ will also lead to a little bit steeper Balmer decrement. However, the anti-correlation trend in relation of $\rm H\alpha/H\beta-\lambda_{\rm Edd}$ is unchanged. The BLR boundary may be also evolved with the accretion rate. If this is the case, the slope of $\rm H\alpha/H\beta-\lambda_{\rm Edd}$ will also slightly change. Furthermore, the covering factor, BLR dynamics and BLR geometry may also change during the changing look in AGNs. Monitoring the shape of broad lines may help to understand this issue, which can shed light on the variations in different lines. In the LOC model we adopted, high density and very low ionization clouds at a large distance from the SMBH were included in our computations, where weakly-ionized gas may emit strong low-ionization features \citep[e.g., Na {\footnotesize I} and Ca {\footnotesize II}, see][]{Ferland1989}. This expectation should be more evident in low continuum states, where the flux of ionizing photons is much reduced and the ionization parameter becomes lower for gas of a given density. To avoid this issue, we also test the LOC model with a distance-dependent density distribution, where we adopt $n\propto r^s$ with $s=-1.5$ \citep[e.g.,][]{Kaspi1999,Rees1989}. We find this model can reproduce our main results (e.g, Figures 2 and 3) but with much weaker neutral emission lines, where the ratio of Ca {\footnotesize II}/H$\beta$ decreases more than one order of magnitude compared to that in the former LOC model. The uncertainties of the LOC model do not affect the main conclusion of anti-correlation of $\rm H\alpha/H\beta-\lambda_{\rm Edd}$ in this work.

\begin{acknowledgments}
We appreciate our referee for very constructive comments and suggestions, which help us a lot to improve our paper. We also thank Luis C. Ho, Jianmin Wang, Xuebing Wu, Xinwu Cao, Junjie Jin, Hengxiao Guo, and Zhicheng He for very helpful discussions on the results, samples, and Cloudy code. This work is supported by the NSFC (grants 12233007, U1931203, U2038107) and the science research grants from the China Manned Space Project (No. CMS-CSST-2021-A06). W.H.Lei. acknowledges support by the National Key R\&D Program of China (Nos. 2020YFC2201400) and by the science research grants from the China Manned Space Project with NO.CMS-CSST-2021-B11. The authors acknowledge Beijng PARATERA Tech CO., Ltd. for providing HPC resources that have contributed to the results reported within this paper.
\end{acknowledgments}

\vspace{5mm}

\software{
          Cloudy \citep{Ferland2017}.
          }
\appendix

\section{Prediction for some emission lines}
\begin{deluxetable*}{lcccc}[h]
\centering
\tabcolsep=0.6cm
\tablecaption{ \textbf{Model predictions for emission-line luminosity.} \label{table2}}
\tablehead{
\colhead{${\rm log}L({\rm line})$} & \colhead{$\dot{m}=1$} & \colhead{$\dot{m}=0.1$} & \colhead{$\dot{m}=0.01$} & \colhead{$\dot{m}=0.001$}
}
\colnumbers
\startdata
$\rm H\alpha$ & 43.66 & 42.80 & 41.96 & 39.81 \\
$L_{5100}-L_{\rm H\alpha}$ observation & 43.55 & 42.74 & 41.88 & 39.55 \\
\hline
$\rm H\beta$ & 43.18 & 42.24 & 41.22 & 38.97 \\
$L_{5100}-L_{\rm H\beta}$ observation & 42.96 & 42.17 & 41.32 & 39.05 \\
\hline
Ly$\alpha$ & 44.57 & 43.54 & 42.23 & 40.00 \\
C {\footnotesize IV}  & 44.04 & 42.82 & 41.95 & 39.91 \\
Mg {\footnotesize II} & 43.36 & 42.46 & 41.76 & 39.84 \\
He {\footnotesize II} & 43.20 & 41.82 & 40.91 & 38.79 \\
\enddata
\tablecomments{The model prediction for luminosities of several typical broad lines for $M_{\rm BH}=10^8 M_\odot$ with different accretion rates. For comparison, we also present the observed line luminosities at give optical continuum luminosity based on $L_{\rm 5100}-L_{\rm H\alpha}$ and $L_{\rm 5100}-L_{\rm H\beta}$ relation  for an AGN sample \citep{Greene2005}. 
}
\end{deluxetable*}

\bibliography{sample631}{}
\bibliographystyle{aasjournal}

\end{document}